%% file: main.tex
\documentclass[letterpaper]{article}
\usepackage[T1]{fontenc}
\usepackage{spconf,amsmath,amssymb,graphicx,url,booktabs,tikz}
\usetikzlibrary{arrows.meta,positioning,calc}
\newif\ifextended
\extendedtrue
\newcommand{\system}{Whisper-Flash}

\input{generated/full-numbers.tex}

\input{generated/throughput-numbers.tex}

\input{generated/sampling-numbers.tex}

\title{Whisper-Flash: Acoustically Conditioned Parallel Drafting\\
for Faster Whisper Decoding}
\input{authors.tex}
\begin{document}
\maketitle

\begin{abstract}
Whisper is a widely used encoder--decoder model for speech recognition. Its encoder reads an utterance in one parallel pass, but its decoder writes the transcript one token at a time, which dominates inference time. Speculative decoding shortens such loops without changing their output: a small drafter guesses several upcoming tokens, and the original model verifies them all in one forward pass. We present \system, a two-layer drafter built on a property of speech recognition: the words still to be written have already been spoken. It reads Whisper's encoded audio and accepted decoder states and proposes eight tokens in a single forward pass. On the complete LibriSpeech test sets, \system\ processes \FullCleanWFSpeed$\times$/\FullOtherWFSpeed$\times$ as much audio per second as greedy decoding with identical outputs, and it remains faster at batch sizes up to 96 and under temperature sampling. Ablations show that direct access to the audio matters most.
\end{abstract}


\section{Introduction}
Whisper~\cite{whisper} is an encoder--decoder Transformer for speech recognition, trained on 680,000 hours of weakly supervised multilingual audio. Its robustness to accents, noise and domain shift has made it a widely used open-source recognizer, so the cost of running it matters at scale.

Whisper transcribes in two stages. The encoder turns up to 30\,s of audio into acoustic representations in one parallel pass; the decoder then writes the transcript one token at a time, running all 32 of its layers for every word piece, punctuation mark and control symbol. The second stage dominates: for representative LibriSpeech requests on an H100, the encoder takes only 5--12\% of the time of greedy decoding.

Each decoding step does little arithmetic. Its cost comes from reading the decoder's weights and running dozens of small kernels, so processing several tokens in one step costs little more than processing one. Speculative decoding~\cite{leviathan,chen} exploits this: a small \emph{drafter} guesses several upcoming tokens, and the original \emph{target} model checks them all in one forward pass, keeping the longest prefix it agrees with and supplying its own token at the first disagreement. The greedy transcript thus stays exactly the target's own, and a rejection rule preserves the target's distribution when sampling, yet each expensive call can now emit several tokens.

How much this helps depends on the drafter, which must guess both accurately and cheaply. Autoregressive drafters such as Distil-Whisper~\cite{distilwhisper} guess well but also write one token at a time, so $K$ guesses cost $K$ sequential drafter steps. Parallel drafters guess a whole block in one forward pass, removing this sequential stage, but each position must then predict its token without seeing the guesses before it.

Speech recognition eases this dilemma in a way text generation cannot. A language model continuing a prompt faces an open future, but the words a recognizer has yet to write have already been spoken and encoded. A parallel drafter can therefore read upcoming tokens from the audio instead of inventing them, provided each proposed position can look at the right part of the recording. The audio cannot, however, reveal how Whisper will write those words or where the transcript currently stands; that knowledge lives in the target decoder's hidden states.

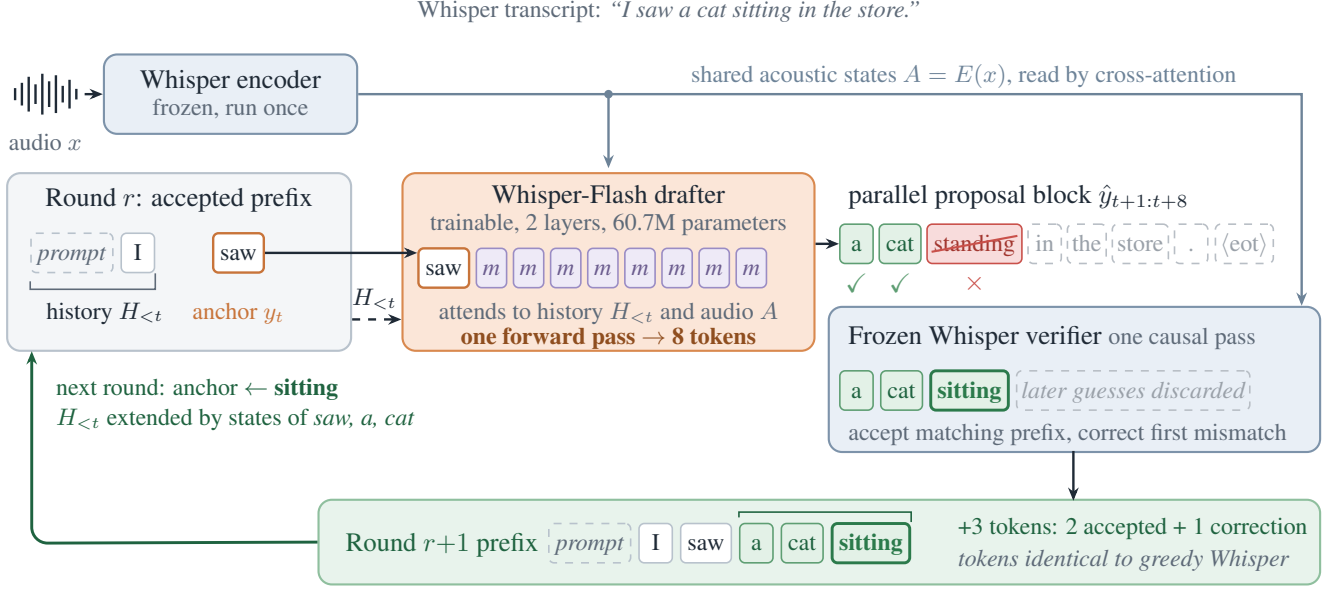
\begin{figure*}[t]
\centering
\input{figures/figure1-9pt.tex}
\caption{\system\ drafts eight tokens in one pass from Whisper's acoustic memory and accepted decoder states. The verifier accepts ``a cat,'' replaces ``standing'' with ``sitting'' and discards later guesses; the correction becomes the next anchor. Words stand in for token IDs.}
\label{fig:method}
\end{figure*}

\system\ combines the two (Fig.~\ref{fig:method}). Following the target-conditioned block drafting of DFlash~\cite{dflash}, a two-layer drafter initialized from Whisper's decoder fills eight mask positions after the latest token. Each position attends to hidden states left by earlier verification passes and cross-attends directly to the encoded audio, so the whole block emerges from one forward pass, while frozen Whisper keeps the final say over every token. We evaluate it on the complete LibriSpeech test sets, across batch sizes and at sampling temperatures, and test what each memory contributes.

\section{Related Work}
Speculative decoding for Whisper has mostly used autoregressive drafters: Distil-Whisper~\cite{distilwhisper} distills the decoder into two layers that share Whisper's encoder, Whisper-Turbo~\cite{whisperturbo} can serve with its own encoder, and WhisperKit~\cite{whisperkit} conditions a recurrent drafter~\cite{redrafter} on Whisper's hidden states. Among parallel drafters, blockwise parallel decoding~\cite{stern} and Medusa~\cite{medusallm} predict several future tokens from the target's current state with separate heads, and Whisper-Medusa~\cite{medusa} applies the idea to Whisper; its block variant adds a cross-attending layer, yet all heads read one vector at the current position. DFlash~\cite{dflash} lets a block of mask positions attend to multi-layer target features, and TreeSpark~\cite{zhou2026treesparkcalibratedloadadaptivedraft} grows calibrated, load-adaptive draft trees from such block drafters; both were designed for language models, whose future is unobserved. Drax~\cite{drax} starts from the audio: its flow-matching recognizer drafts for Whisper in two function evaluations without the target's states. \system\ gives every proposed position its own query into both the target's history and the audio.

\section{Method}
\subsection{Propose, verify and advance}
\label{sec:verify}
Let $x$ be the audio and $y_{\leq t}$ the transcript emitted so far; greedy Whisper would continue with
\begin{equation}
 y_{t+1}=\arg\max_v p_{\mathrm{target}}(v\mid x,y_{\leq t}).
\end{equation}
\system\ instead proposes $K$ candidates $\hat y_{t+1:t+K}$ at once. The latest emitted token $y_t$, not yet processed by the target, serves as the \emph{anchor}. One causal target pass over the anchor and the candidates yields the target's prediction at every position. The verifier keeps candidates while they agree and emits the target's token at the first disagreement, or a bonus token if all $K$ agree; in Fig.~\ref{fig:method}, one call thus advances the transcript by three tokens. The correction or bonus becomes the next anchor, and only the old anchor and accepted candidates enter the target cache and the drafter's history, so the greedy output is unchanged; we confirm this token by token, including punctuation and end-of-transcript (EOS).

For sampling at temperature $T>0$, the drafter samples each candidate from its distribution $q_i$, and the verifier accepts it with probability $\min(1,p_i(\hat y_i)/q_i(\hat y_i))$, where $p_i$ is the target distribution under the same temperature and token rules. At the first rejection it resamples from the normalized residual $(p_i-q_i)_+$, and after a fully accepted block it samples a bonus, which preserves the target's distribution in exact arithmetic~\cite{leviathan,chen}.

\subsection{A drafter that reads the transcript and the audio}
The drafter draws on two memories Whisper has already computed. The acoustic memory $A=E(x)$ is the encoder output. The history memory $H_{<t}$ holds the target's decoder states at accepted positions before the anchor, taken from five blocks, concatenated and projected to the drafter's width; since verification produces them anyway, the drafter summarizes the prefix at no extra cost. A round's input is the anchor embedding followed by $K=8$ copies of a learned mask vector $m$,
\begin{equation}
 X_t=[\operatorname{Emb}(y_t),m,\ldots,m],
\end{equation}
and each mask position predicts its own token. In every drafter layer, the block's queries attend to the history and to the block itself within one softmax, bidirectionally inside the block, with rotary embeddings (RoPE)~\cite{rope} marking text positions; cross-attention then lets each position read the acoustic memory, followed by a feed-forward update.

The target's states do carry acoustic information, but they were computed to predict only the next token, so positions further ahead need their own path to the audio. Initialized from Whisper's first two decoder blocks, the drafter adds 60.7M trainable parameters, while the encoder, target decoder, embeddings and output head stay frozen and shared: a small network learns to anticipate the recognizer, not to replace it.

\subsection{Training and execution}
The drafter is trained to predict Whisper's own greedy continuation rather than the reference, since only agreement with the target counts. Cached runs of frozen Whisper supply these tokens and hidden states. Each utterance contributes up to 16 packed anchor blocks; each block sees the audio but only history before its own anchor, never another block. With $v_{bj}$ masking padding and labels after EOS, we minimize
\begin{equation}
 \mathcal L=-\frac{\sum_{b,j}v_{bj}e^{-(j-1)/4}\log q_\theta(y_{t_b+j})}
 {\sum_{b,j}v_{bj}e^{-(j-1)/4}},\quad j=1,\ldots,8.
\end{equation}
The decaying weights favor early positions, because a later guess counts only if all earlier ones are accepted; checkpoints are selected by mean accepted-prefix length on validation data. At inference, CUDA graphs~\cite{cudagraphs} replay prefill, drafting and verification. Greedy decoding runs through the same graph-captured target, so our speedups come from emitting more tokens per call, not from removing launch overhead.

\section{Experimental Setup}
The target is Whisper large-v3 in FP16, transcribing known-English speech without timestamps. The drafter is trained on the 94.5\,h of LibriSpeech train-clean-100~\cite{librispeech} and then continued on a 1,126.6\,h pool from LibriSpeech, TED-LIUM~3~\cite{tedlium3} and English VoxPopuli~\cite{voxpopuli}, with speaker-disjoint validation; this broader pool sped up TED-LIUM and VoxPopuli development audio by 17\% and 22\% over the train-clean-100 drafter. The architecture and block size were fixed before testing. The baseline is greedy decoding through the same graph-captured target; batched comparisons add Hugging Face SDPA and faster-whisper's batched core~\cite{fasterwhisper}.

All timings use one H100 80GB GPU, two warm-up runs and three timed repeats. The complete test sets (2,620 test-clean and 2,939 test-other utterances) are decoded one request at a time in rotating method order, with the 16 recordings longer than 30\,s split into independent 30\,s segments of at most 432 tokens. These runs are timed from GPU log-mel features to the last token; batched studies time the full path from waveform to text. Neither includes file I/O, loading or graph capture. Throughput is given in seconds of audio processed per second (audio s/s) and generated text tokens per second (tok/s); $S_G$ relates it to greedy decoding on the same audio and batch size, with confidence intervals from 2,000 paired bootstrap resamples. $\alpha$ is the fraction of proposals accepted and $\tau$ the tokens emitted per target call. Outputs are compared as token IDs, including EOS.

\section{Results}
\subsection{Complete test sets}
On the complete test sets, \system\ processes \ThroughputCleanWF\ and \ThroughputOtherWF\ seconds of audio per second on test-clean and test-other, \FullCleanWFSpeed$\times$ and \FullOtherWFSpeed$\times$ the \ThroughputCleanGraph\ and \ThroughputOtherGraph\ of greedy decoding (95\% CI [\FullCleanWFCILow, \FullCleanWFCIHigh] and [\FullOtherWFCILow, \FullOtherWFCIHigh]). All 5,559 transcripts match the greedy tokens in every repeat (WER 1.98\%/3.83\%). Because Whisper encodes a fixed 30\,s window whatever the duration, the encoder's cost, which speculation cannot reduce, weighs most on short utterances: the speedup grows from 2.28$\times$ below 5\,s to 4.02$\times$ at 20--30\,s.

\subsection{Throughput across batch sizes}
\begin{table}[t]
\centering\small
\setlength{\tabcolsep}{4pt}
\begin{tabular}{rrrrrr}
\toprule
 & \multicolumn{2}{c}{Greedy decoding} & \multicolumn{2}{c}{\textbf{\system}} & \\
\cmidrule(lr){2-3}\cmidrule(lr){4-5}
Batch & tok/s & audio s/s & tok/s & audio s/s & $S_G$ \\
\midrule
\multicolumn{6}{l}{\emph{64 development clips (427.2\,s)}} \\
1 & 181.7 & 51.93 & 530.8 & 151.68 & 2.92 \\
2 & 266.7 & 76.22 & 765.4 & 218.71 & 2.87 \\
4 & 397.8 & 113.69 & 1052.0 & 300.64 & 2.64 \\
8 & 554.8 & 158.54 & 1270.4 & 363.04 & 2.29 \\
16 & 680.3 & 194.42 & 1413.3 & 403.88 & 2.08 \\
32 & 809.7 & 231.37 & 1577.5 & 450.80 & 1.95 \\
\midrule
\multicolumn{6}{l}{\emph{192 development clips (1,286.7\,s)}} \\
32 & 851.2 & 249.98 & 1586.7 & 466.00 & 1.86 \\
64 & 972.5 & 285.63 & 1686.1 & 495.18 & 1.73 \\
96 & 1027.8 & 301.87 & 1695.5 & 497.97 & 1.65 \\
\bottomrule
\end{tabular}
\caption{Throughput by batch size on an H100, timed from waveform to text: generated text tokens (tok/s) and seconds of audio processed per second. $S_G$ is the speedup over greedy decoding at the same batch size, whose tokens \system\ matches at every batch size.}
\label{tab:batch}
\end{table}

\begin{figure}[t]
\centering
\includegraphics{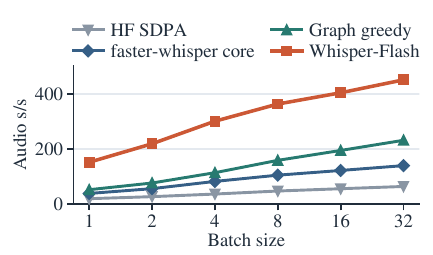}
\caption{Throughput versus batch size on the 64 development clips of Table~\ref{tab:batch}, timed from waveform to text.}
\label{fig:batch}
\end{figure}

Batching keeps the GPU busier and leaves speculation less idle capacity to exploit, so we measure throughput from batch size 1 to 96 (Table~\ref{tab:batch}). On 64 development clips, \system\ rises from 530.8 to 1577.5 tokens and from 151.68 to 450.80 seconds of audio per second between batch sizes 1 and 32, while its lead over greedy decoding narrows from 2.92$\times$ to 1.95$\times$. Figure~\ref{fig:batch} adds two common engines: at batch size 32, faster-whisper processes 139.37\,s of audio per second, with its own feature frontend changing some transcripts, and Hugging Face SDPA 63.70. On a separate 192-clip set, \system\ levels off near 500 audio s/s at batch sizes 64 and 96, still 1.73$\times$ and 1.65$\times$ faster than greedy decoding; at batch size 96, the encoder already takes about half of its time. These are static offline batches.

\subsection{Comparison with Whisper-Medusa}
\begin{table}[t]
\centering\small
\setlength{\tabcolsep}{4pt}
\begin{tabular}{lrr}
\toprule
Engine & Audio s/s $\uparrow$ & WER (\%) $\downarrow$ \\
\midrule
\multicolumn{3}{l}{\emph{Whisper large-v2 systems}} \\
HF greedy & 24.72 & 2.90 \\
HF + length penalty & 24.84 & 2.90 \\
Whisper-Medusa Linear & 35.82 & 5.80 \\
Whisper-Medusa Block & 32.14 & 3.79 \\
Whisper-Medusa Block$^\dagger$ & 29.70 & 2.90 \\
\midrule
\multicolumn{3}{l}{\emph{Whisper large-v3 systems}} \\
HF SDPA & 19.26 & \textbf{2.18} \\
Graph greedy & 51.93 & \textbf{2.18} \\
\textbf{\system} & \textbf{151.68} & \textbf{2.18} \\
\bottomrule
\end{tabular}
\caption{Released Whisper-Medusa checkpoints and \system\ on the 64 clips of Table~\ref{tab:batch} (batch size one, waveform to text). Medusa uses large-v2, Hugging Face Transformers and its official acceptance rule; $\dagger$ is our exact strict-argmax Medusa-Block.}
\label{tab:medusa}
\end{table}

Whisper-Medusa~\cite{medusa} is the closest prior parallel drafter for Whisper. Its released checkpoints build on large-v2 and run in Hugging Face Transformers, so Table~\ref{tab:medusa} compares each system with its own greedy baseline. Medusa's official rule gains 1.30--1.45$\times$ but raises WER from 2.90\% to 3.79--5.80\%; kept exact, Medusa-Block gains only 1.20$\times$, whereas \system\ gains 2.92$\times$ with identical tokens.

\subsection{Why both memories matter}
\label{sec:ablation}
\begin{table}[t]
\centering\small
\setlength{\tabcolsep}{3.5pt}
\begin{tabular}{lrrrr}
\toprule
Conditioning & $\alpha$ (\%) & $\tau$ & Audio s/s & $S_G$ \\
\midrule
\textbf{Audio + history} & \textbf{47.37} & \textbf{4.60} & \textbf{157.07} & \textbf{2.81} \\
Without direct audio & 16.48 & 2.23 & 97.55 & 1.69 \\
Without target history & 30.15 & 3.29 & 128.63 & 2.30 \\
\bottomrule
\end{tabular}
\caption{Conditioning ablation on 128 development clips, batch size one. The drafters share initialization, data, seed and budget (60.7M/47.5M/52.5M trainable parameters). Each drafter ran in its own job, and $S_G$ is relative to greedy decoding timed in that job (55.9/57.7/55.9 audio s/s); all outputs match greedy decoding.}
\label{tab:conditioning}
\end{table}

Table~\ref{tab:conditioning} isolates the two memories with three drafters retrained on train-clean-100 under identical conditions. With both, 47.37\% of the proposals are accepted and throughput is 2.81$\times$ that of greedy decoding. Without the target history, acceptance drops to 30.15\% and the speedup to 2.30$\times$; without direct access to the audio, acceptance collapses to 16.48\% and the speedup to 1.69$\times$, although the history still carries the target's audio-informed states. This variant resembles DFlash's target-only conditioning, and its gap to the full drafter supports our premise that a parallel drafter for speech must look ahead in the audio.

\subsection{Sampling at a fixed temperature}
\begin{table}[t]
\centering\small
\setlength{\tabcolsep}{3.5pt}
\begin{tabular}{rrrrrr}
\toprule
 & \multicolumn{2}{c}{Audio s/s} & & \multicolumn{2}{c}{WER (\%)} \\
$T$ & AR & \textbf{WF} & Gain & AR & \textbf{WF} \\
\midrule
0 & 56.26 & 173.68 & 3.09 & 2.24 & 2.24 \\
0.2 & 55.06 & 166.18 & 3.02 & 2.21 & 2.21 \\
0.4 & 55.22 & 165.84 & 3.00 & 2.14 & 2.33 \\
0.6 & 55.34 & 164.51 & 2.97 & 2.35 & 2.37 \\
0.8 & 55.49 & 159.12 & 2.87 & 2.77 & 2.95 \\
1 & 55.22 & 147.59 & 2.67 & 5.57 & 5.29 \\
\bottomrule
\end{tabular}
\caption{Temperature sampling on 256 development clips (batch size one, three runs per clip). AR samples autoregressively from the target; gain is \system\ (WF) over AR.}
\label{tab:temperature}
\end{table}

Table~\ref{tab:temperature} uses the rejection rule of Sec.~\ref{sec:verify} with the unchanged drafter on 256 development clips at temperatures from 0 to 1, with top-$p=1$ and no top-$k$ or fallback. \system\ stays \SamplingMinGain--\SamplingMaxGain$\times$ faster than autoregressive sampling from the same target, and the gain erodes only gradually as acceptance falls from 57.5\% at $T=0$ to 44.8\% at $T=1$. WER differences between the two reflect sampling variation.

\section{Conclusion}
In speech recognition, the tokens still to be written are already present in the encoded audio. \system\ exploits this with a small drafter that reads the audio alongside Whisper's decoder states, roughly tripling greedy throughput on LibriSpeech without changing any output token and keeping most of the gain under sampling. Other domains, languages, long-form audio and beam search remain to be evaluated.

\clearpage
\section{Compliance with Ethical Standards}
This research study was conducted retrospectively using human subject data made available in open access by LibriSpeech~\cite{librispeech}, TED-LIUM~3~\cite{tedlium3} and VoxPopuli~\cite{voxpopuli}. Ethical approval was not required as confirmed by the licenses attached with the open access data.

\section{Acknowledgments}
This work was supported by Boson AI, which provided the computing resources. Huapeng Zhou is an employee of Boson AI. The other authors have no relevant financial or non-financial interests to disclose.

\bibliographystyle{IEEEbib}
\bibliography{references,dataset-references,appendix-references,revision-references}
\ifextended
\clearpage
\appendix
\setcounter{table}{0}\renewcommand{\thetable}{A\arabic{table}}
\setcounter{figure}{0}\renewcommand{\thefigure}{A\arabic{figure}}
\input{appendix-compact.tex}
\fi
\end{document}

%% file: generated/full-numbers.tex
\newcommand{\FullCleanWFSpeed}{3.16}
\newcommand{\FullCleanWFCILow}{3.13}
\newcommand{\FullCleanWFCIHigh}{3.20}

\newcommand{\FullOtherWFSpeed}{2.85}
\newcommand{\FullOtherWFCILow}{2.82}
\newcommand{\FullOtherWFCIHigh}{2.88}

%% file: generated/throughput-numbers.tex
\newcommand{\ThroughputCleanGraph}{60.98}
\newcommand{\ThroughputOtherGraph}{58.56}

\newcommand{\ThroughputCleanWF}{192.80}
\newcommand{\ThroughputOtherWF}{166.91}

%% file: generated/sampling-numbers.tex
\newcommand{\SamplingMinGain}{2.67}
\newcommand{\SamplingMaxGain}{3.09}

%% file: authors.tex
\name{Huapeng Zhou$^{1,\ast}$ \quad Huayu Wang$^{2}$ \quad
Junkai Wu$^{2}$ \quad Kangqi Wang$^{3}$ \quad Xinyu Wang$^{4,\ast}$}
\address{$^{1}$Boson AI \quad $^{2}$University of Washington\\
$^{3}$Independent Researcher \quad $^{4}$McGill University\\
$^{\ast}$Corresponding authors. Contact: Huapeng Zhou \texttt{<zhouhp.me@gmail.com>}}

%% file: figures/figure1-9pt.tex
\definecolor{wfInk}{HTML}{1F2A36}
\definecolor{wfMuted}{HTML}{5E6A78}
\definecolor{wfWire}{HTML}{6D8398}
\definecolor{wfEncFill}{HTML}{E9EFF6}
\definecolor{wfEncLine}{HTML}{7D93AA}
\definecolor{wfPreFill}{HTML}{F4F6F8}
\definecolor{wfPreLine}{HTML}{B8C2CC}
\definecolor{wfDrFill}{HTML}{FBE6D8}
\definecolor{wfDrLine}{HTML}{D9894F}
\definecolor{wfAnchor}{HTML}{C8743A}
\definecolor{wfBrown}{HTML}{8F4E1F}
\definecolor{wfMaskFill}{HTML}{EEEAF7}
\definecolor{wfMaskLine}{HTML}{9A8CC4}
\definecolor{wfMaskText}{HTML}{6B5C8E}
\definecolor{wfAccFill}{HTML}{E3F1E7}
\definecolor{wfAccLine}{HTML}{5E9E73}
\definecolor{wfAccText}{HTML}{1F6340}
\definecolor{wfCorr}{HTML}{2B7A4B}
\definecolor{wfRejFill}{HTML}{F8DCDC}
\definecolor{wfRejLine}{HTML}{C9504A}
\definecolor{wfRejText}{HTML}{A2362F}
\definecolor{wfDisc}{HTML}{B8BEC5}
\definecolor{wfDiscText}{HTML}{9AA1A9}
\definecolor{wfNextFill}{HTML}{E8F3EC}
\definecolor{wfNextLine}{HTML}{8FBF9E}
\begin{tikzpicture}[
  font=\small, text=wfInk, >={Stealth[length=5pt,width=4pt]},
  panel/.style={draw, rounded corners=5pt, line width=0.8pt},
  tok/.style={draw, rounded corners=2pt, line width=0.6pt, inner xsep=2.6pt,
    inner ysep=0pt, minimum height=0.52cm, minimum width=0.44cm, anchor=west,
    text depth=0.06cm, text height=0.25cm},
  plain/.style={tok, draw=wfPreLine, fill=white},
  prompt/.style={tok, draw=wfPreLine, dashed, text=wfMuted, font=\small\itshape},
  anchortok/.style={tok, draw=wfAnchor, line width=0.9pt, fill=white},
  mask/.style={tok, draw=wfMaskLine, fill=wfMaskFill, text=wfMaskText,
    font=\small\itshape, inner xsep=1.2pt, minimum width=0.40cm},
  acc/.style={tok, draw=wfAccLine, fill=wfAccFill, text=wfAccText},
  corr/.style={tok, draw=wfCorr, line width=1.1pt, fill=wfAccFill, text=wfCorr,
    font=\small\bfseries},
  rej/.style={tok, draw=wfRejLine, fill=wfRejFill, text=wfRejText},
  disc/.style={tok, draw=wfDisc, dashed, text=wfDiscText, inner xsep=1.8pt},
  wire/.style={draw=wfWire, line width=0.9pt},
  lbl/.style={font=\small, text=wfMuted},
]
\node[lbl, anchor=base] at (8.85, 7.62)
  {Whisper transcript: \emph{``I saw a cat sitting in the store.''}};
\foreach \x/\h in {0.20/0.16,0.29/0.30,0.38/0.46,0.47/0.26,0.56/0.54,0.65/0.34,
                   0.74/0.50,0.83/0.22,0.92/0.38,1.01/0.14}
  \draw[wfInk, line width=0.9pt] (\x, 6.60-\h/2) -- (\x, 6.60+\h/2);
\node[lbl, anchor=north] at (0.60, 6.20) {audio $x$};
\node[panel, draw=wfEncLine, fill=wfEncFill, minimum width=3.35cm,
      minimum height=1.05cm, align=center] (enc) at (3.05, 6.60)
  {{\normalsize Whisper encoder}\\[-1pt]{\color{wfMuted}frozen, run once}};
\draw[->, wfInk, line width=0.9pt] (1.12, 6.60) -- (enc.west);
\draw[wire, ->] (enc.east) -- (17.22, 6.60) -- (17.22, 3.80);
\fill[wfWire] (8.05, 6.60) circle (1.6pt);
\draw[wire, ->] (8.05, 6.60) -- (8.05, 5.62);
\node[lbl, text=wfWire, anchor=base] at (12.75, 6.76)
  {shared acoustic states $A=E(x)$, read by cross-attention};
\node[panel, draw=wfPreLine, fill=wfPreFill, minimum width=4.55cm,
      minimum height=2.35cm] (pre) at (2.37, 4.38) {};
\node[anchor=base] at (2.37, 5.12) {\normalsize Round $r$: accepted prefix};
\node[prompt] (p0) at (0.40, 4.50) {prompt};
\node[plain, right=0.09cm of p0] (p1) {I};
\node[anchortok, right=0.75cm of p1] (p2) {saw};
\draw[wfMuted, line width=0.6pt] ([yshift=-3pt]p0.south west) -- ++(0,-3pt)
  -| ([yshift=-3pt]p1.south east) -- ++(0,3pt);
\node[lbl, text=wfInk, anchor=base] at ([yshift=-17pt]$(p0.south)!0.5!(p1.south)$)
  {history $H_{<t}$};
\node[anchor=base, text=wfAnchor] at ([yshift=-17pt]p2.south) {anchor $y_t$};
\node[panel, draw=wfDrLine, fill=wfDrFill, minimum width=5.45cm,
      minimum height=2.35cm] (dr) at (8.05, 4.38) {};
\node[anchor=base] at (8.05, 5.16) {\normalsize Whisper-Flash drafter};
\node[lbl, anchor=base] at (8.05, 4.80) {trainable, 2 layers, 60.7M parameters};
\node[anchortok] (d0) at (5.52, 4.30) {saw};
\foreach \i [evaluate=\i as \j using int(\i-1)] in {1,...,8}
  \node[mask, right=0.07cm of d\j] (d\i) {m};
\node[lbl, anchor=base] at (8.05, 3.62) {attends to history $H_{<t}$ and audio $A$};
\node[anchor=base, text=wfBrown, font=\small\bfseries] at (8.05, 3.30)
  {one forward pass $\rightarrow$ 8 tokens};
\draw[->, wfInk, line width=0.9pt] (p2.east) -- (d0.west |- p2.east);
\draw[->, wfInk, line width=0.9pt, dashed] (pre.east |- 0,3.62) --
  node[above, font=\small, pos=0.45] {$H_{<t}$} (dr.west |- 0,3.62);
\node[anchor=base west] at (11.10, 5.16) {\normalsize parallel proposal block $\hat y_{t+1:t+8}$};
\node[acc] (q1) at (11.10, 4.62) {a};
\node[acc, right=0.06cm of q1] (q2) {cat};
\node[rej, right=0.06cm of q2] (q3) {standing};
\draw[wfRejLine, line width=0.8pt] ([xshift=2pt,yshift=-3pt]q3.west) -- ([xshift=-2pt,yshift=3pt]q3.east);
\node[disc, right=0.06cm of q3] (q4) {in};
\node[disc, right=0.06cm of q4] (q5) {the};
\node[disc, right=0.06cm of q5] (q6) {store};
\node[disc, right=0.06cm of q6] (q7) {.};
\node[disc, right=0.08cm of q7, font=\small] (q8) {$\langle$eot$\rangle$};
\draw[->, wfInk, line width=0.9pt] (dr.east |- q1.west) -- (q1.west);
\node[text=wfAccLine, font=\normalsize\bfseries] at ([yshift=-7pt]q1.south) {\checkmark};
\node[text=wfAccLine, font=\normalsize\bfseries] at ([yshift=-7pt]q2.south) {\checkmark};
\node[text=wfRejLine, font=\normalsize\bfseries] at ([yshift=-7pt]q3.south) {$\times$};
\node[panel, draw=wfEncLine, fill=wfEncFill, minimum width=6.50cm,
      minimum height=1.92cm, anchor=north west] (ver) at (10.95, 3.80) {};
\node[anchor=base west] at (11.10, 3.30)
  {{\normalsize Frozen Whisper verifier}\;{\color{wfMuted}one causal pass}};
\node[acc] (v1) at (11.10, 2.68) {a};
\node[acc, right=0.08cm of v1] (v2) {cat};
\node[corr, right=0.08cm of v2] (v3) {sitting};
\node[disc, right=0.08cm of v3, font=\small\itshape] (v4) {later guesses discarded};
\node[lbl, anchor=base west] at (11.10, 2.02)
  {accept matching prefix, correct first mismatch};
\node[panel, draw=wfNextLine, fill=wfNextFill, minimum width=13.25cm,
      minimum height=1.12cm, anchor=south west] (nxt) at (4.20, 0.10) {};
\draw[->, wfInk, line width=0.9pt] (14.20, 1.88) -- (14.20, 1.22);
\node[anchor=base west, text=wfAccText] at (4.45, 0.55) {\normalsize Round $r{+}1$ prefix};
\node[prompt] (n0) at (7.25, 0.66) {prompt};
\node[plain, right=0.09cm of n0] (n1) {I};
\node[plain, right=0.09cm of n1] (n2) {saw};
\node[acc, right=0.09cm of n2] (n3) {a};
\node[acc, right=0.09cm of n3] (n4) {cat};
\node[corr, right=0.09cm of n4] (n5) {sitting};
\draw[wfAccText, line width=0.6pt] ([yshift=2pt]n3.north west) -- ++(0,2pt)
  -| ([yshift=2pt]n5.north east) -- ++(0,-2pt);
\node[anchor=base west, text=wfAccText] at (12.55, 0.78) {+3 tokens: 2 accepted + 1 correction};
\node[anchor=base west, text=wfMuted, font=\small\itshape] at (12.55, 0.36)
  {tokens identical to greedy Whisper};
\draw[->, wfAccText, line width=1.1pt, rounded corners=6pt]
  (nxt.west) -- (0.42, 0.66) -- (0.42, 3.20);
\node[anchor=base west, text=wfAccText] at (0.62, 2.62)
  {next round: anchor $\leftarrow$ \textbf{sitting}};
\node[anchor=base west, text=wfAccText] at (0.62, 2.22)
  {$H_{<t}$ extended by states of \emph{saw, a, cat}};
\end{tikzpicture}

%% file: appendix-compact.tex
\section{Implementation details}
\label{app:impl}
\subsection{Anchor, cache and output budget}
The prefill pass processes the language and task prompt and emits the first transcript token, which becomes the first anchor, so the emitted transcript always runs one token ahead of the target cache. Before a round with anchor $y_t$, the target key/value cache and the drafter history hold exactly the positions before $t$. If the cache has length $L$ and the first $a$ candidates match, rollback keeps $L+1+a$ positions, the old anchor plus the accepted candidates, and the correction or bonus becomes the next unprocessed anchor. An accepted EOS ends the transcript without a bonus. With $R$ output tokens left, at most $R-1$ candidates are proposed, reserving one slot for the target's own token.

\subsection{Drafter architecture}
History features are the residual streams after target decoder blocks 1, 8, 15, 22 and 29 (zero-based), taken before the final layer normalization; their 6,400-dimensional concatenation is projected to the drafter width of 1,280. The projected history and the current block form the keys and values of one attention softmax, so a masked position sees all accepted history, the anchor and the other masked positions, but never history at or beyond its anchor. RoPE uses base 10,000, FP32 phases and explicit text positions, so cached history keys keep their indices across rounds. The drafter has its own acoustic key/value projections, while the encoder states, token embedding and output head are shared with the frozen target.

\subsection{Training}
Labels are frozen Whisper's greedy continuations rather than reference transcripts, extracted with the history features in one teacher-forced pass; trajectories that reach the length limit or have fewer than two tokens are excluded. Each utterance contributes up to 16 anchors sampled without replacement, and packed blocks attend only to history strictly before their own anchor. The drafter is first trained on train-clean-100 (26,749 utterances, 94.5\,h, 235 speakers) for 40,000 updates with AdamW, batch size 4, peak learning rate $6\times10^{-4}$, 4\% warmup, cosine decay and gradient clipping at 1.0, using FP32 parameters with BF16 autocast; validation on 1,790 utterances from 16 held-out speakers selects step 38,000. It is then continued for 80,000 updates at peak learning rate $2\times10^{-4}$ on a pool of 371,235 clips (1,126.6\,h), sampling LibriSpeech clean, LibriSpeech other, TED-LIUM~3 and VoxPopuli with probabilities 0.3, 0.5, 0.1 and 0.1; the run sees 656.5\,h of unique audio. The ablation drafters of Table~\ref{tab:conditioning} use only the first stage, with identical seeds and anchor draws; their speedups have 95\% confidence intervals of [2.64, 2.97], [1.63, 1.74] and [2.20, 2.40].

\subsection{Sampling}
For temperature sampling, draft and target logits receive the same token suppression and temperature scaling, and softmax and acceptance probabilities are computed in FP32 while both models run in FP16. Each masked position is sampled independently from its draft distribution given the history and anchor, which keeps the proposal probabilities that the rejection rule needs. EOS and the output budget are handled as in greedy decoding. This implements fixed-temperature sampling, not Whisper's temperature fallback, beam search or best-of selection.

\subsection{Execution and output equality}
CUDA graphs are captured once for prefill, drafting and each needed verification length, with preallocated caches and tensor-valued positions; rejected positions stay allocated but masked. Every request re-encodes its audio and refreshes the acoustic key/value projections of target and drafter. Output equality is measured against greedy decoding through the same graph-captured target. Compared with eager PyTorch decoding, this graph-captured greedy decoding differs on two sentence-final punctuation tokens across 512 development clips, caused by FP16 argmax ties, with identical normalized WER; Whisper-Flash reproduces the graph-captured tokens in every case.

\section{Additional results}
\label{app:results}
\subsection{Throughput of all engines}
\begin{table}[tb]
\centering\small
\setlength{\tabcolsep}{3.5pt}
\begin{tabular}{rrrrr}
\toprule
Batch & HF SDPA & FW core & Greedy & \textbf{\system} \\
\midrule
\multicolumn{5}{l}{\emph{64 development clips}} \\
1 & 19.26 & 38.33 & 51.93 & \textbf{151.68} \\
2 & 26.66 & 55.72 & 76.22 & \textbf{218.71} \\
4 & 36.15 & 82.02 & 113.69 & \textbf{300.64} \\
8 & 47.13 & 104.71 & 158.54 & \textbf{363.04} \\
16 & 55.46 & 121.77 & 194.42 & \textbf{403.88} \\
32 & 63.70 & 139.37 & 231.37 & \textbf{450.80} \\
\midrule
\multicolumn{5}{l}{\emph{192 development clips}} \\
32 & 63.89 & 141.26 & 249.98 & \textbf{466.00} \\
64 & 64.31 & 156.59 & 285.63 & \textbf{495.18} \\
96 & 63.82 & 160.91 & 301.87 & \textbf{497.97} \\
\bottomrule
\end{tabular}
\caption{Audio s/s of all engines by batch size (H100, waveform to text), the data behind Fig.~\ref{fig:batch} and Table~\ref{tab:batch}. FW denotes faster-whisper.}
\label{app:tab:engines}
\end{table}

Table~\ref{app:tab:engines} lists every engine at every batch size. Peak PyTorch memory of \system\ is 14.9, 26.4 and 38.0\,GiB at batch sizes 32, 64 and 96. faster-whisper computes its own log-mel features: on the 192 clips its WER is 2.39\% at batch sizes 32 and 64 and 2.87\% at 96, against 2.20\% for greedy decoding and \system, and feeding it Whisper's own features reduces the clips whose normalized text differs from greedy decoding from 20/20/22 to 1/1/3 (WER 2.17/2.17/2.66\%).

\subsection{Where the time goes}
\begin{table}[tb]
\centering\small
\setlength{\tabcolsep}{3pt}
\begin{tabular}{rlrr}
\toprule
Batch & Request or group & Greedy (\%) & \textbf{\system} (\%) \\
\midrule
1 & p25 length & 12.01 & 34.19 \\
1 & p50 length & 8.51 & 20.94 \\
1 & p90 length & 4.94 & 15.30 \\
32 & group 0 & 29.25 & 50.01 \\
32 & group 5 & 29.60 & 55.82 \\
96 & group 0 & 30.23 & 52.04 \\
96 & group 1 & 37.14 & 57.59 \\
\bottomrule
\end{tabular}
\caption{Encoder share of request time, from CUDA events that include host submission gaps. Batch one uses representative requests at the 25th, 50th and 90th length percentiles; larger batches use complete groups.}
\label{app:tab:profile}
\end{table}

Removing most decoder calls makes the fixed encoder cost more visible (Table~\ref{app:tab:profile}): at batch size one the encoder takes 5--12\% of greedy decoding but 15--34\% of \system, and at batch sizes 32 and 96 it reaches half of \system's time. In the two batch-96 groups, verification calls fall from 70 and 48 with greedy decoding to 15 and 10, so further gains at large batch sizes must come from the encoder rather than the decoder.

\subsection{Speedup by utterance length}
\begin{table}[tb]
\centering\small
\setlength{\tabcolsep}{3.5pt}
\begin{tabular}{lrr}
\toprule
Duration (s) & Utterances & Speedup \\
\midrule
$[0,5)$ & 2508 & 2.278 \\
$[5,10)$ & 1948 & 3.045 \\
$[10,20)$ & 965 & 3.807 \\
$[20,30)$ & 122 & 4.024 \\
$[30,\infty)$ & 16 & 3.280 \\
\bottomrule
\end{tabular}
\caption{Speedup over greedy decoding on the complete test sets by utterance duration (test-clean and test-other combined, batch size one). Recordings longer than 30\,s are decoded as independent 30\,s segments.}
\label{app:tab:strata}
\end{table}

The fixed encoder pass also explains the length dependence in Table~\ref{app:tab:strata}: short utterances have few tokens to accelerate, while 20--30\,s utterances reach a 4.02$\times$ speedup. Utterance durations have medians of 5.79 and 5.15\,s on test-clean and test-other, and outputs have medians of 20 and 18 tokens.

\subsection{Design choices}
\begin{table}[tb]
\centering\small
\setlength{\tabcolsep}{3pt}
\begin{tabular}{llrrr}
\toprule
Layers & Position & $\alpha$ (\%) & $a$ & ms \\
\midrule
\multicolumn{5}{l}{\emph{1,914 clips, 6,000 updates}} \\
2 & Abs. & 9.68 & 0.77 & 97.4 \\
2 & RoPE & 31.94 & 2.56 & 59.3 \\
\multicolumn{5}{l}{\emph{train-clean-100, 40,000 updates}} \\
2 & Abs. & 48.12 & 3.85 & 47.1 \\
5 & Abs. & 52.72 & 4.22 & 47.8 \\
\textbf{2} & \textbf{RoPE} & \textbf{56.58} & \textbf{4.53} & \textbf{44.7} \\
5 & RoPE & 58.93 & 4.71 & 47.6 \\
\bottomrule
\end{tabular}
\caption{Position encoding and depth, trained with matched data, seed and budget; 32 dev-clean clips, batch size one. $a$ is the mean number of accepted candidates per round.}
\label{app:tab:position}
\end{table}

\begin{table}[tb]
\centering\small
\setlength{\tabcolsep}{3.5pt}
\begin{tabular}{rrrrr}
\toprule
$K$ & ms & $S_G$ & $\alpha$ (\%) & $\tau$ \\
\midrule
1 & 84.74 & 1.41 & 92.54 & 1.89 \\
2 & 66.37 & 1.80 & 87.14 & 2.67 \\
4 & 49.94 & 2.39 & 74.74 & 3.85 \\
\textbf{8} & 43.77 & 2.72 & 54.76 & 5.17 \\
\bottomrule
\end{tabular}
\caption{Verifying only the first $K$ of the eight proposals of the same drafter; 64 development clips, batch size one, timed from waveform to text.}
\label{app:tab:budget}
\end{table}

Table~\ref{app:tab:position} shows why the drafter uses two layers with RoPE. With little data, RoPE is essential (31.9\% vs.\ 9.7\% acceptance); with the full train-clean-100 split, absolute positions recover much of the gap, but RoPE still leads. Five layers accept slightly more proposals but cost more drafting time than they save. Table~\ref{app:tab:budget} shows that later positions are accepted less often, yet verifying all eight proposals still lowers latency, because each extra proposal is almost free to check; the speedup at $K=8$ has a 95\% confidence interval of [2.54, 2.90].

\subsection{Training data}
\begin{table}[tb]
\centering\small
\setlength{\tabcolsep}{3.5pt}
\begin{tabular}{lrrrr}
\toprule
Continuation & First (\%) & $\alpha$ (\%) & $\tau$ & ms \\
\midrule
No continuation & 82.02 & 46.45 & 4.52 & 41.53 \\
train-clean-100 only & 81.42 & 46.75 & 4.54 & 41.52 \\
LibriSpeech (925.6\,h) & 87.69 & 53.95 & 5.09 & 38.84 \\
\textbf{+ TED-LIUM, VoxPopuli} & 87.24 & 54.21 & 5.12 & 38.77 \\
\bottomrule
\end{tabular}
\caption{Continuing the train-clean-100 drafter for 80,000 updates on different pools, evaluated on 128 dev-other clips at batch size one. ``First'' is first-candidate acceptance; the last row is the released drafter.}
\label{app:tab:continuation}
\end{table}

Continuing on train-clean-100 alone changes little, while adding the other LibriSpeech splits raises acceptance and speed on dev-other (Table~\ref{app:tab:continuation}); relative to the drafter without continuation, the released drafter is 1.07$\times$ faster there (95\% CI [1.05, 1.10]). Adding TED-LIUM and VoxPopuli keeps the LibriSpeech gains, and relative to the drafter without continuation it speeds up the TED-LIUM and VoxPopuli development sets by 17\% and 22\%.

\subsection{Temperature sampling}
\begin{table}[tb]
\centering\small
\setlength{\tabcolsep}{3pt}
\input{tables/temperature-details.tex}
\caption{Temperature-sampling diagnostics for Table~\ref{tab:temperature}. Gain CI is the 95\% interval of the throughput gain over autoregressive sampling; $\Delta$WER is \system\ minus autoregressive sampling in percentage points, with paired 95\% intervals.}
\label{app:tab:temperature}
\end{table}
Acceptance falls slowly up to $T=0.6$ and faster beyond it, and every gain interval stays above 2.5$\times$ (Table~\ref{app:tab:temperature}). The WER intervals are pointwise and unadjusted across the six temperatures; only $T=0.4$ excludes zero (+0.18 points). Each clip is sampled three times with fixed seeds, 2,000 paired bootstrap resamples are stratified by split, and no request reaches the 432-token limit.

%% file: tables/temperature-details.tex
\begin{tabular}{rrrrr}
\toprule
$T$ & $\alpha$ (\%) & $\tau$ & Gain CI & $\Delta$WER [CI] \\
\midrule
0 & 57.53 & 5.39 & [2.99, 3.19] & 0.00 [0.00, 0.00] \\
0.2 & 57.20 & 5.37 & [2.93, 3.11] & 0.00 [-0.10, +0.09] \\
0.4 & 56.48 & 5.31 & [2.91, 3.10] & +0.18 [+0.03, +0.33] \\
0.6 & 55.46 & 5.23 & [2.88, 3.06] & +0.01 [-0.15, +0.17] \\
0.8 & 51.66 & 4.94 & [2.78, 2.96] & +0.18 [-0.20, +0.59] \\
1 & 44.79 & 4.42 & [2.58, 2.77] & -0.28 [-1.25, +0.61] \\
\bottomrule
\end{tabular}

%% file: main.bbl
\begin{thebibliography}{10}

\bibitem{whisper}
Alec Radford, Jong~Wook Kim, Tao Xu, Greg Brockman, Christine McLeavey, and
  Ilya Sutskever,
\newblock ``Robust speech recognition via large-scale weak supervision,''
\newblock in {\em Proc. ICML}, 2023, pp. 28492--28518.

\bibitem{leviathan}
Yaniv Leviathan, Matan Kalman, and Yossi Matias,
\newblock ``Fast inference from transformers via speculative decoding,''
\newblock in {\em Proc. ICML}, 2023, pp. 19274--19286.

\bibitem{chen}
Charlie Chen, Sebastian Borgeaud, Geoffrey Irving, Jean-Baptiste Lespiau,
  Laurent Sifre, and John Jumper,
\newblock ``Accelerating large language model decoding with speculative
  sampling,''
\newblock {\em arXiv preprint arXiv:2302.01318}, 2023.

\bibitem{distilwhisper}
Sanchit Gandhi, Patrick von Platen, and Alexander~M. Rush,
\newblock ``{Distil-Whisper}: Robust knowledge distillation via large-scale
  pseudo labelling,''
\newblock {\em arXiv preprint arXiv:2311.00430}, 2023.

\bibitem{dflash}
Jian Chen, Yesheng Liang, and Zhijian Liu,
\newblock ``{DFlash}: Block diffusion for flash speculative decoding,''
\newblock in {\em Proc. ICML}, 2026.

\bibitem{whisperturbo}
{OpenAI},
\newblock ``{Whisper large-v3-turbo},''
  \url{https://huggingface.co/openai/whisper-large-v3-turbo}, 2024.

\bibitem{whisperkit}
Atila Orhon, Arda Okan, Berkin Durmus, Zach Nagengast, and Eduardo Pacheco,
\newblock ``{WhisperKit}: On-device real-time {ASR} with billion-scale
  transformers,''
\newblock in {\em Proc. ICML Workshop on On-Device Learning for Foundational
  Models}, 2025.

\bibitem{redrafter}
Yunfei Cheng, Aonan Zhang, Xuanyu Zhang, Chong Wang, and Yi~Wang,
\newblock ``Recurrent drafter for fast speculative decoding in large language
  models,''
\newblock {\em arXiv preprint arXiv:2403.09919}, 2024.

\bibitem{stern}
Mitchell Stern, Noam Shazeer, and Jakob Uszkoreit,
\newblock ``Blockwise parallel decoding for deep autoregressive models,''
\newblock in {\em Proc. NeurIPS}, 2018.

\bibitem{medusallm}
Tianle Cai, Yuhong Li, Zhengyang Geng, Hongwu Peng, Jason~D. Lee, Deming Chen,
  and Tri Dao,
\newblock ``{Medusa}: Simple {LLM} inference acceleration framework with
  multiple decoding heads,''
\newblock in {\em Proc. ICML}, 2024.

\bibitem{medusa}
Yael Segal-Feldman, Aviv Shamsian, Aviv Navon, Gill Hetz, and Joseph Keshet,
\newblock ``Whisper in {Medusa}'s ear: Multi-head efficient decoding for
  transformer-based {ASR},''
\newblock in {\em Proc. ICASSP}, 2025.

\bibitem{zhou2026treesparkcalibratedloadadaptivedraft}
Huapeng Zhou, Huayu Wang, and Xinyu Wang,
\newblock ``{TreeSpark}: Calibrated, load-adaptive draft trees for
  semi-autoregressive speculative decoding,'' arXiv preprint arXiv:2609.22098,
  2026.

\bibitem{drax}
Aviv Navon, Aviv Shamsian, Neta Glazer, Yael Segal-Feldman, Gill Hetz, Joseph
  Keshet, and Ethan Fetaya,
\newblock ``{Drax}: Speech recognition with discrete flow matching,''
\newblock {\em arXiv preprint arXiv:2510.04162}, 2025.

\bibitem{rope}
Jianlin Su, Yu~Lu, Shengfeng Pan, Ahmed Murtadha, Bo~Wen, and Yunfeng Liu,
\newblock ``{RoFormer}: Enhanced transformer with rotary position embedding,''
\newblock {\em arXiv preprint arXiv:2104.09864}, 2021.

\bibitem{cudagraphs}
{PyTorch Contributors},
\newblock ``{CUDA} semantics: {CUDA} graphs,'' {PyTorch} 2.7 documentation,
\newblock \url{https://docs.pytorch.org/docs/2.7/notes/cuda.html}, accessed
  September 21, 2026.

\bibitem{librispeech}
Vassil Panayotov, Guoguo Chen, Daniel Povey, and Sanjeev Khudanpur,
\newblock ``{LibriSpeech}: An {ASR} corpus based on public domain audio
  books,''
\newblock in {\em Proc. ICASSP}, 2015.

\bibitem{tedlium3}
Fran\c{c}ois Hernandez, Vincent Nguyen, Sahar Ghannay, Natalia Tomashenko, and
  Yannick Est\`eve,
\newblock ``{TED-LIUM 3}: Twice as much data and corpus repartition for
  experiments on speaker adaptation,''
\newblock in {\em Proc. SPECOM}, 2018, pp. 198--208.

\bibitem{voxpopuli}
Changhan Wang, Morgane Riviere, Ann Lee, Anne Wu, Chaitanya Talnikar, Daniel
  Haziza, Mary Williamson, Juan Pino, and Emmanuel Dupoux,
\newblock ``{VoxPopuli}: A large-scale multilingual speech corpus for
  representation learning, semi-supervised learning and interpretation,''
\newblock in {\em Proc. ACL-IJCNLP}, 2021, pp. 993--1003.

\bibitem{fasterwhisper}
{SYSTRAN},
\newblock ``Faster {Whisper} transcription with {CTranslate2},''
  \url{https://github.com/SYSTRAN/faster-whisper},
\newblock Accessed September 20, 2026.

\end{thebibliography}
